# Robust Voxelization and Visualization by Improved Tetrahedral Mesh Generation

Joseph Chen
Graduate Institute of
Networking and Multimedia,
National Taiwan University
h111030nehs@gmail.com

Ko-Wei Tai
Department of Computer
Science and Information
Engineering,
National Taiwan University
b08902080@ntu.edu.tw

Wen-Chin Chen
Department of Computer
Science and Information
Engineering,
National Taiwan University
wcchen@csie.ntu.edu.tw

Ming Ouhyoung
Department of Computer
Science and Information
Engineering,
National Taiwan University
ming@csie.ntu.edu.tw

*Abstract*—When obtaining interior 3D voxel data from triangular meshes, most existing methods fail to handle low quality meshes which happen to take up a big portion on the internet. In this work we present a robust voxelization method that is based on tetrahedral mesh generation within a user defined error bound. Compared to other tetrahedral mesh generation methods, our method produces much higher quality tetrahedral meshes as the intermediate outcome, which allows us to utilize a faster voxelization algorithm that is based on a stronger assumption. We show the results compared to various methods including the state-of-the-art.

Our contribution includes a framework which takes triangular mesh as an input and produces voxelized data, a proof to an unproved algorithm that performs better than the state-of-the-art, and various experiments including parallelization built on the GPU and CPU. We further tested our method on various datasets including Princeton ModelNet and Thingi10k to show the robustness of the framework, where near 100% availability is achieved, while others can only achieve around 50%.

*Keywords*—voxelization, computation geometry, tetrahedral mesh, mesh reconstruction

1. INTRODUCTION

Just as pixels are important in 2D graphics, voxel representation plays an important role in 3D graphics. It has welcoming natures such as the ease to perform binary operations and is particularly ideal for simulations when per-element simulation is needed. Voxelized data can be further categorized into solid and surface data. The solid binary voxel representation, which this work is going to focus on, is arguably the most accurate and intuitive data structure for 3D models(Figure 1). However, due to its sparse nature, most 3D models are not stored in voxel forms for its storage cost is too expensive.

Figure 1. Solid binary voxel representation of a surface mesh model

The demand on solid voxelized data has grew tremendous these days with techniques such as 3D printing and 3D Convolutional Neural Networks (3DCNN), which is growing popular at a thundering pace. In comparison of generating their own voxelized data, alternatively, researchers often gather triangular surface meshes and voxelize them using a pipeline. However, a big portion of triangular mesh data in the internet isn't produced by professionals and includes flaws making current data gathering pipeline tools easy to stall.

Researchers has been performing this pipelining process by first generating tetrahedral mesh to obtain space information, then voxelizing the tetrahedral mesh according to whether the sampling point is within a tetrahedral or not. This method is precise and can somewhat solve the flawed mesh problem depending on the tetrahedral mesh generating algorithm used. In other words, a nicely voxelized solid model heavily relies on a nice tetrahedral mesh of that model.

In this paper we present a robust framework that can successfully voxelize nearly 100% of our testing data. Furthermore, our work voxelizes tetrahedral mesh faster than other methods on the assumption that tetrahedral mesh will have nice quality. Detailed experiment and results are given in §5.

While surveying for related works, we noticed that [1] has recently presented a work with similar scope and impressive outcomes. It integrated with the GPU through GLSL, making the work independent from GPU vendors. However, as they mentioned in their framework, the model needs to be topological closed for their algorithm to work. In the rest of the context, we consider their work as the state of the art so far, and we will cover other related papers in the following section.

## 2. RELATED WORK

*Overview*

Related work of this paper can be briefly divided into three fields of research. Voxelization methods regards all work that are popular to this work, which includes surface and solid voxelization methods, as well as dense and sparse voxelization methods. Mesh repair techniques holds an important part in our work, making it possible to handle imperfect meshes, which is the backbone of the robustness issue in our system. Tetrahedral mesh generation processes the surface triangle mesh into solid meshes containing volume information. It also somewhat involves in both mesh repair and voxelization.

*Voxelization*

Voxelization methods, that dates back to [2], has different focuses along its evolution due to hardware capabilities. Early in the 1990s, research focused on providing accurate and robust results. The attention had moved on to time efficient processing since [3]. Various work pioneered by works of [4], [5] has pushed the field further as a popular research topic for computer graphics.

The amount of different approaches to achieve this task is remarkable. Starting from conventional approaches such as those presented in [6] [7] [8], and [9], [10] presents a scan plane/line method for both solid and surface voxelization. [4] uses the slicemap algorithm from their previous work [11]. (An algorithm different from [12]'s sliceMap for medical rasterizations.)

Forest, Barthe [13] presented a surface voxelization using octrees that runs in real time. [14] presents an irregular hexahedral structure, and Fang and Chen [15] presents one of the most cited work using slice-based rasterization in the standard rendering pipeline.

The most related studies from our work includes [16] and [1], using tetrahedral mesh rasterization which successfully produced better outcomes than that of other methods. [1], proposed a parallelized method that takes no assumption of the GPU vendor running on the target machine.

*Mesh Reconstruction*

The eagerness to obtain a better model has existed ever since 3D models are created. The attention has only grown bigger and bigger after the invention of 3D printers. [17] introduced an algorithm to match non-overlapping curves. [18] and [19] fetched information from overlapping fragments as additional restrictions to enhance the quality of the result.

Some newer researches include machine learning approaches [20] [21]. These methods are fast and robust, but may produce noise which is fatal for voxelizing. [22] employed an algorithm especially for bone models, using an AABB structure to reach efficiency. There is another work on biological tissues, [23] furthermore examined into the materials used in 3D printing to mimic living organisms. [24] achieved superior results by taking account of a user defined ε-envelope while transforming the surface mesh into a solid tetrahedral mesh, which is ideal for voxelization and will be introduced in the next section.

*Tetrahedral Mesh generation*

To voxelize, or 3D rasterize, over generated tetrahedral mesh is by far the most intuitive and accurate method to perform voxelization. The state of the art method [1] also uses this approach. The tetrahedral mesh generated by their method connects every triangle facet to the calculated centroid of the model, which naturally produces faulty tetrahedral meshes and is undesired in models with more faces. Better tetrahedral mesh generating techniques include Delaunay triangulation and background grid insertions.

The most studied method of tetrahedral mesh generation is to perform a 3D Delaunay triangulation. Due to its accuracy and well optimized structure, it is also the most widely used method in commercial software. [25-31]

Grid insertion is another popular method on tetrahedral mesh generation, and the same concept is also used in voxelization. Grid insertion is intuitive in both tetrahedral mesh generation and voxelization. The to be processed model is inserted into a grid of neat, tidy shapes (either tetrahedral or cubes), aligned by its bounding box.

These algorithms produce good results in the interior of meshes but suffer from accuracy and quality around the boundary. It is also computational expensive when handling sparse models. However, by combining this idea with other methods like Delaunay triangulation, we are able to create quality mesh while maintaining the process speed.

## 3. METHOD

This work is inspired by the works [32] [33], which we will give a brief review in the upcoming sections. The outline of our proposed method is illustrated in Figure 2.

Figure 2. Outline of our proposed framework

*Tetrahedral Mesh generating*

In order to obtain a valid tetrahedral mesh, we adapted the method of [32], which is based on a background mesh insertion with mesh improvement processing both before and after the insertion process.

This tetrahedral mesh generation algorithm requires a user defined ε as an input. ε indicates the user's tolerance towards errors. An obvious artifact that the paper also demonstrates suggests that the same ε also decides the amount of zigzag edges the algorithm will produce when dealing with straight lines (Figure 3).

We found out that the error in setting ε to as much as 1e-2, which is 10 times the default value, still isn't large enough to produce a major difference when voxelized. Furthermore, slight difference produced by setting the error bound larger can be efficiently used as a data augmentation technique for 3DCNNs.

Figure 3. Zigzag artifact when generating tetrahedral meshes.
From left to right are, respectively, the ground truth, 2e-3, 5e-3 and 1e-2 error bounds

*Voxelization*

Also known as 3D rasterization, this process is made to be implemented closely related to the types of GPU. By the tetrahedral mesh generation step we previously went through, we do not need to rely on watertightness or even topologically closed. Furthermore, since we have tetrahedral mesh with a quality exceeding previous works, we adapt a faster algorithm that assumes the tetrahedral mesh to have a large minimum angle, no overlapping, and with small variance.

We iterate over every bounding box of the tetrahedral and apply the fastest point-in-tetrahedral test algorithm accordingly. Phillip B's algorithm was invented in June 2020 but was left unproven in correctnessuntil the point this paper is written. As a part of the proof of the robustness, we would show our proof in §4.

*Parallelization*

We implemented our parallelization with OpenMP and CUDA. Since loading the data into GPU memory takes up resources, voxelizing models with a smaller tetrahedral count in GPU isn't guaranteed to be significantly faster. We pick a few models from the thingi10k dataset that has an order difference of tetrahedral count to illustrate the overhead in GPU memory copying (Table 1).

| **Vertices** | **Triangles** | **Tetrahedral** | **Nodes** | **CPU** | **CPU+GPU** |
|---|---|---|---|---|---|
| $1.3*10^6$ | $2.7*10^6$ | $10^7$ | $1.3*10^6$ | 98.931s | 9.77s |
| $4.8*10^4$ | $10^5$ | $10^6$ | $2.7*10^5$ | 2.947s | 2.776s |
| $10^4$ | $2*10^4$ | $10^5$ | $2.7*10^4$ | 0.503s | 1.665s |

Table 1. Comparison of GPU parallelization. The models are voxelized into a $256^3$ binary 3D grid.

# 4. PHILIPP B'S ALGORITHM

Philipp B's algorithm plays a crucial role in sampling voxel points from tetrahedral meshes. It was first brought up by Philipp Beisel as an answer to Denis Lolik's question: " How to check whether the point is in the tetrahedron or not?"[34]. Although it was measured to be faster than all other algorithms with the same purpose, the correctness of the algorithm was left unproven.

We will start the proof by first defining the theorem.

*Theorem 1*

Let there be arbitrary points *P, A, B, C, D* in 3D Cartesian coordinates and $\mathbf{a} = \overrightarrow{PA}$, $\mathbf{b} = \overrightarrow{PB}$, $\mathbf{c} = \overrightarrow{PC}$, $\mathbf{d} = \overrightarrow{PD}$

Point *P* lies in the interior of the tetrahedral spanned by the four arbitrary points *ABCD* if and only if one of the two formula holds:

$$D_a > 0 \text{ and } D_b < 0 \text{ and } D_c > 0 \text{ and } D_d < 0 \quad or$$
$$D_a < 0 \text{ and } D_b > 0 \text{ and } D_c < 0 \text{ and } D_d > 0$$

Where,

$D_a$ denotes the determinant of $V_a = \{\mathbf{b}, \mathbf{c}, \mathbf{d}\}$, the matrix whose row vectors are **b, c, d,**
$D_b$ denotes the determinant of $V_b = \{\mathbf{a}, \mathbf{c}, \mathbf{d}\}$, the matrix whose row vectors are **a, c, d,**
$D_c$ denotes the determinant of $V_c = \{\mathbf{a}, \mathbf{b}, \mathbf{d}\}$, the matrix whose row vectors are **a, b, d,**
$D_d$ denotes the determinant of $V_d = \{\mathbf{a}, \mathbf{b}, \mathbf{c}\}$, the matrix whose row vectors are **a, b, c.**

*Proof*

We are going to complete the proof step by step through four lemmas.

*Lemma 1.1*

There are 2 kinds of tetrahedral with the same vertices' names but opposite signed volumes.

*Proof*

Figure 4. Type I and Type II tetrahedral

Given a tetrahedral *ABCD*, consider triangle *ABC* viewing from the opposite direction of point *D* in regarding plane *ABC*. In other words, consider triangle *ABC* from the outer side of the tetrahedral. We define the tetrahedral *type I* if the vertices read *ABC* in counterclockwise order, and *type II* is the vertices reads *ACB* in counterclockwise order, as in Fig 4.

Since point D decides the orientation of triangle ABC, it is clear that Type I and Type II tetrahedral differ from each other. Additionally, since there is not a third way to represent triangle ABC and that point D decided the orientation, all tetrahedral lie in one of the two types provided. It is left to prove that they have opposite signed values.

Figure 5. Signed volume of a tetrahedral

Recall that a signed volume of a tetrahedral can be calculated by one sixth of the determinant of the three vectors that span the tetrahedral. Using the tetrahedral from Figure 1, we can get the following equations:

$$Type \ I \ volume \ = \ \frac{\det(\overrightarrow{AB} \ \overrightarrow{AC} \ \overrightarrow{AD})}{6}$$
$$Type \ II \ volume \ = \ \frac{\det(\overrightarrow{AC} \ \overrightarrow{AB} \ \overrightarrow{AD})}{6}$$

It is clear that the two volumes can be obtained by exchanging two rows in the determinant of the other volume, which leads to an opposite signed volume.

*Lemma 1.2*

If *P* lies in tetrahedral *ABCD*, then

$D_a > 0$ and $D_b < 0$ and $D_c > 0$ and $D_d < 0$, or
$D_a < 0$ and $D_b > 0$ and $D_c < 0$ and $D_d > 0$

where  $D_a$ denotes the determinant of $V_a = \{\mathbf{b, c, d}\}$, the matrix whose row vectors are **b, c, d**,
$D_b$ denotes the determinant of $V_b = \{\mathbf{a, c, d}\}$, the matrix whose row vectors are **a, c, d**,
$D_c$ denotes the determinant of $V_c = \{\mathbf{a, b, d}\}$, the matrix whose row vectors are **a, b, d**,
$D_d$ denotes the determinant of $V_d = \{\mathbf{a, b, c}\}$, the matrix whose row vectors are **a, b, c**,
holds.

*Proof*

We will prove this lemma by going through all possible outcomes. Recall that the signed volume of a tetrahedral is determined by the right-hand rule. Since we have only two types of tetrahedral, we will go through the both of them and check if the lemma holds.

By applying the right-hand rule to each tetrahedral *PABC*, *PACD*, *PABD*, and *PBCD*, we get the following results.

Figure 6. Signed volume results for each sub-tetrahedral

*Lemma 2.1*

If $D_a > 0$ and $D_b < 0$ and $D_c > 0$ and $D_d < 0$,
or $D_a < 0$ and $D_b > 0$ and $D_c < 0$ and $D_d > 0$

Holds, the plane *P* and any 2 vertices form divide the other 2 into different sides of the plane.

*Proof*
1. $D_d = \det(\mathbf{abc}) = -\det(\mathbf{bac}) = \det(\mathbf{bca})$

   $D_a = \det(\mathbf{bcd})$,

   We know that $D_d$ and $D_a$ have different signs, meaning that **a** and **d** must point towards different sides of the plane **b** and **c** spans, by the right-hand rule. Thus, gives us the fact that *A* and *D* are divided by plane *PBC*. Similar processes can be applied to other point sets.

2. $D_b = \det(\mathbf{acd}) = -\det(\mathbf{adc})$

   $D_c = \det(\mathbf{abd}) = -\det(\mathbf{adb})$.

Since $D_b$ and $D_c$ have differ signs, *B* and *C* are divided by plane *PAD* by the right-hand rule.

3. $D_a = \det(\mathbf{bcd}) = -\det(\mathbf{cbd}) = \det(\mathbf{cdb})$

   $D_b = \det(\mathbf{acd}) = -\det(\mathbf{cad}) = \det(\mathbf{cda})$

Since $D_a$ and $D_b$ have differ signs, *A* and *B* are divided by plane *PCD* by the right-hand rule.

4. $D_a = \det(\mathbf{bcd}) = -\det(\mathbf{bdc})$

$$D_c = \det(\mathbf{abd}) = -\det(\mathbf{bad}) = \det(\mathbf{bda}).$$

Since $D_a$ and $D_c$ share the same signs, det(**bdc**) and det(**bda**) have differ signs. Thus, *A* and *C* are divided by plane *PBD* by the right-hand rule.

5. $D_b = \det(\mathbf{acd})$

   $D_d = \det(\mathbf{abc}) = -\det(\mathbf{acb}).$

Since $D_b$ and $D_d$ share the same signs, det(**acd**) and det(**acb**) have differ signs. Thus, *B* and *D* are divided by plane *PAC* by the right-hand rule.

6. $D_c = \det(\mathbf{abd})$

   $D_d = \det(\mathbf{abc}).$

Since $D_c$ and $D_d$ have differ signs, *C* and *D* are divided by plane *PAB* by the right-hand rule.

Since all cases are proven correct, the plane *P* and any 2 vertices form divide the other 2 into different sides of the plane.

*Lemma 2.2*

If P is not in the interior of ABCD, then there must be two vertices on the same side of which P and the other 2 vertices form.

*Proof*

Let *P* be a point outside of *ABCD*, we know that tetrahedral *PABC*, *PABD*, *PACD*, and *PBCD* must then have overlapping volumes. Without loss of generality, assume *PABC* and *PABD* have overlap volumes.

Consider plane *PAB*, point *C* and *D* must be on the same side of *PAB* to overlap each other, thus completing the proof.

From the four above lemmas, we complete the only if side by lemma 1.1 and 1.2, the if side from lemma 2.1 and 2.2.

## 5. EXPERIMENTS AND RESULTS

In this section we will illustrate some experimental results we accomplished using the presented work. We also compared our method to other algorithms to accentuate the robustness and capability of pipelining. Main experiments are based on an Intel i7-10700 CPU with 16GB memory, and an NVidia RTX 3080 with

Figure 7. Visualization results by our framework

10GB memory. The input/output operations are done on a Micron MX500 SSD.

We compared our results with [1]. Table 2 shows the benchmark performance proposed by their paper. Each model is voxelized into 5 million samples, with a resolution of $256^3$. The entire framework is implemented in C++, running on Ubuntu 18.04.1, Intel Xeon CPU E5-2650, and an NVidia RTX8000. For testing the performance of the storage input and output, we used two RAID0 HDDs.

|                      | Ours       | Ogayar-Anguita et al. |
|----------------------|------------|-----------------------|
| GPU Voxelization     | 0.102778h  | 21.02h                |
| *GPU-CPU transfers*  | 18.07917h  | 16.28h                |
| *GPU memory allocation* | 412.0569h | 462.5h             |

Table 2. Results on voxelizing into 5 million samples with an output resolution of $256^3$

*Testing Datasets*

We tested our method on various popular CAD models, including Princeton ModelNet40 and ModelNet10, also Thingi10k by [35] for testing on non-manifold meshes. Figure shows some of our results. We compared our tetrahedral mesh generation method with the most widely used commercial software Tetgen[36]. 3000 models in Thingi10k is tested and the timeout is set to 20 minutes. Only if the tetrahedral mesh is generated within 20 minutes it would be further voxelized and verified as a successful voxelization.

|                        | Ours  | Tetgen |
|------------------------|-------|--------|
| Timeout (20m)          | 10    | 46     |
| Fail to produce outcome| 8     | 1183   |
| Overall success rate   | 99.4% | 59.1%  |

Table 3. Rate of successfully producing voxelized outcome tested on 3000 thingi10K models, using different tetrahedral mesh generation methods

*Error Bounds*

The ε is a controllable parameter in the tetrahedral mesh generation algorithm. However, it has a big chance containing zigzags within the error bound when dealing to long straight lines/planes. This error can be smoothed out by the voxelization sampling process if the error bound is small enough. We experimented how small the error should be without major loss in. In the thingi10k dataset, we observed that the voxelization result doesn't differ much when ε is less than 1e-2, and nearly identical results to the ground truth is produced when ε is less than 1e-3.

*Octrees*

Due to its sparse nature, voxels are often calculated and stored in octree data formats. Since the tetrahedral in our work are relatively regular compared to existing methods, we generally have a good bounding box volume to tetrahedral volume ratio, which results in only a handful sample candidates. Even without the octree structure, our algorithm iterates over every tetrahedral faster than other algorithms iterating over each sample point.

6. CONCLUSION, DISCUSSION AND FUTURE WORK

In this paper we presented a robust 3D surface mesh voxelization method that is suit for pipelining over dirty meshes found in the wild. It handles even non-manifold, non-watertight models. We also provided a proof to a point in tetrahedral algorithm that had its correctness unproved. As an addition to the above contributions, we also included parallel specs and experiment results for various tweaks.

On the downside of the method, this two-pass structure leads to a result slower than the state of the art right now. Currently we complete the tetrahedral mesh generation part before performing our voxel sampling. We believe that there is space for optimization by combining the background grid insertion method in tetrahedral mesh generation with the grid method for voxelization sampling. Transferring our algorithm to a GLSL based parallelization, excluding the GPU vendor constraint is also a future work of ours.

We also have application ideas about connecting to 3DCNN and data augmenting, but consider them out of the scope of this paper.

COMPLIANCE WITH ETHICAL STANDARDS

*Conflict of Interest*: The authors declare that they have no conflict of interest.
*Ethical approval*: This chapter does not contain any studies with human participants or animals performed by any of the authors.
*Informed consent*: Informed consent was obtained from all individual participants included in the study.